\documentstyle[graphicx,pifont,amssymb,amsmath]{mn}

\title[]{Chandra observations of the nearby spiral galaxy NGC6503}

\author[Lira etal.]
{P.~Lira,$^{1,2}$  R.~Johnson,$^3$ A.~Lawrence,$^4$\\
$^1$ Departamento de Astronom\'{\i}a, Universidad de Chile, Casilla 36-D, Santiago, Chile\\
$^2$ Department of Physics \& Astronomy, University of Leicester, Leicester LE1 7RH, UK\\
$^3$ European Southern Observatory, Casilla 19001, Santiago 19, Chile\\
$^3$ Institute for Astronomy, University of Edinburgh, Royal
Observatory, Blackford Hill, Edinburgh EH9 3HJ, Scotland\\
} 

\begin{document}

\maketitle

\begin{abstract} 

We have observed the nearby spiral galaxy NGC6503 using Chandra. Seven
discrete sources associated with the galaxy have been found, one of
them coincident with its Liner-starburst nucleus. One of the sources
corresponds to a ULX with L$_{\rm x} \ga 10^{39}$ ergs s$^{-1}$.
Previous ROSAT observations of the galaxy show that this source has
varied by at least a factor 3 in the last 6 years. No evidence is
found for strong diffuse emission in the nuclear region or the
presence of a low luminosity AGN.

\end{abstract}

Galaxies: individual: NGC6503--X-rays:galaxies

\section{Introduction}

NGC6503 is a low luminosity ($M_{B} = -18.28$) edge-on Sc galaxy at a
distance of 7Mpc. It has been the subject of several kinematic studies
which show an unexpected drop in the stellar velocity dispersion in
the inner region of the galaxy (Bottema \& Gerritsen 1997). Based on
the large observed [NII]/H$\alpha$ and [SII]/H$\alpha$ line ratios Ho,
Filippenko \& Sargent (1995) classified the nucleus as a
transition-Seyfert-2 nucleus. With a narrow H$\alpha$ luminosity $\sim
4\times 10^{37}$ ergs s$^{-1}$ this would be one of the lowest
luminosity Seyfert nucleus known.

However, diagnostic diagrams using the [OI]$\lambda 6300$,
[OII]$\lambda 3717$ and [OIII]$\lambda 5007$ optical lines has shown
that the nuclear activity in NGC6503 is better classified as
borderline between starburst and LINER (Lira etal 2002) and therefore
the presence of a low luminosity AGN has been put into dispute.

ROSAT HRI X-ray observations of NGC6503 were reported by Lira etal
(2000). The image of the galaxy showed an extended nuclear source of
luminosity $\sim 10^{39}$ ergs s$^{-1}$ in the 0.5-2.4 keV energy
range. The presence of this source, together with the early
Seyfert-like classification of the nucleus, made this galaxy a
interesting target for high resolution Chandra observations, which
would enable us to study the low luminosity active nucleus and the
circumnuclear region in detail.

In this paper, we show that the X-ray Chandra observations confirm
that NGC6503 is an example of the activity seen in normal spiral
galaxies harbouring a moderate starburst nucleus, while no evidence
for the presence of an AGN is found.


\begin{figure*}
\centering
\begin{minipage}[c]{0.7\textwidth}
\centering
\begin{Huge}
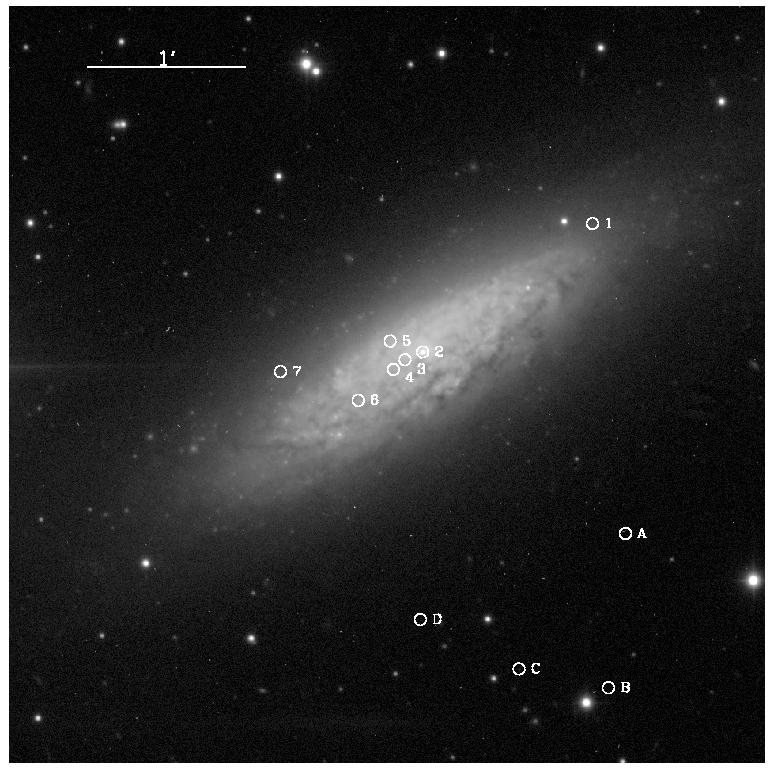
\end{Huge}
\end{minipage}%
\begin{minipage}[c]{0.3\textwidth}
\centering 
\includegraphics[angle=90,scale=0.25]{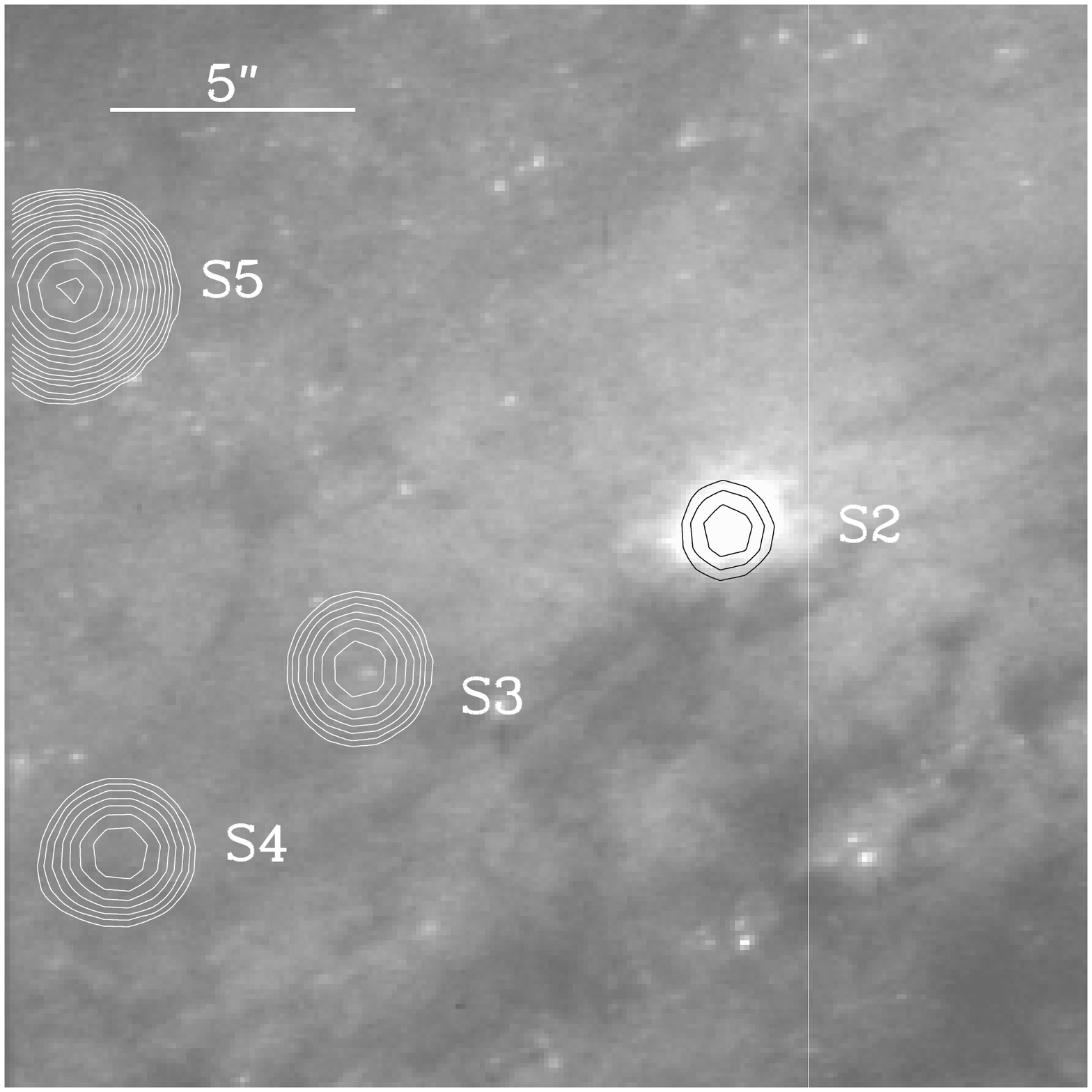}
\includegraphics[scale=0.25]{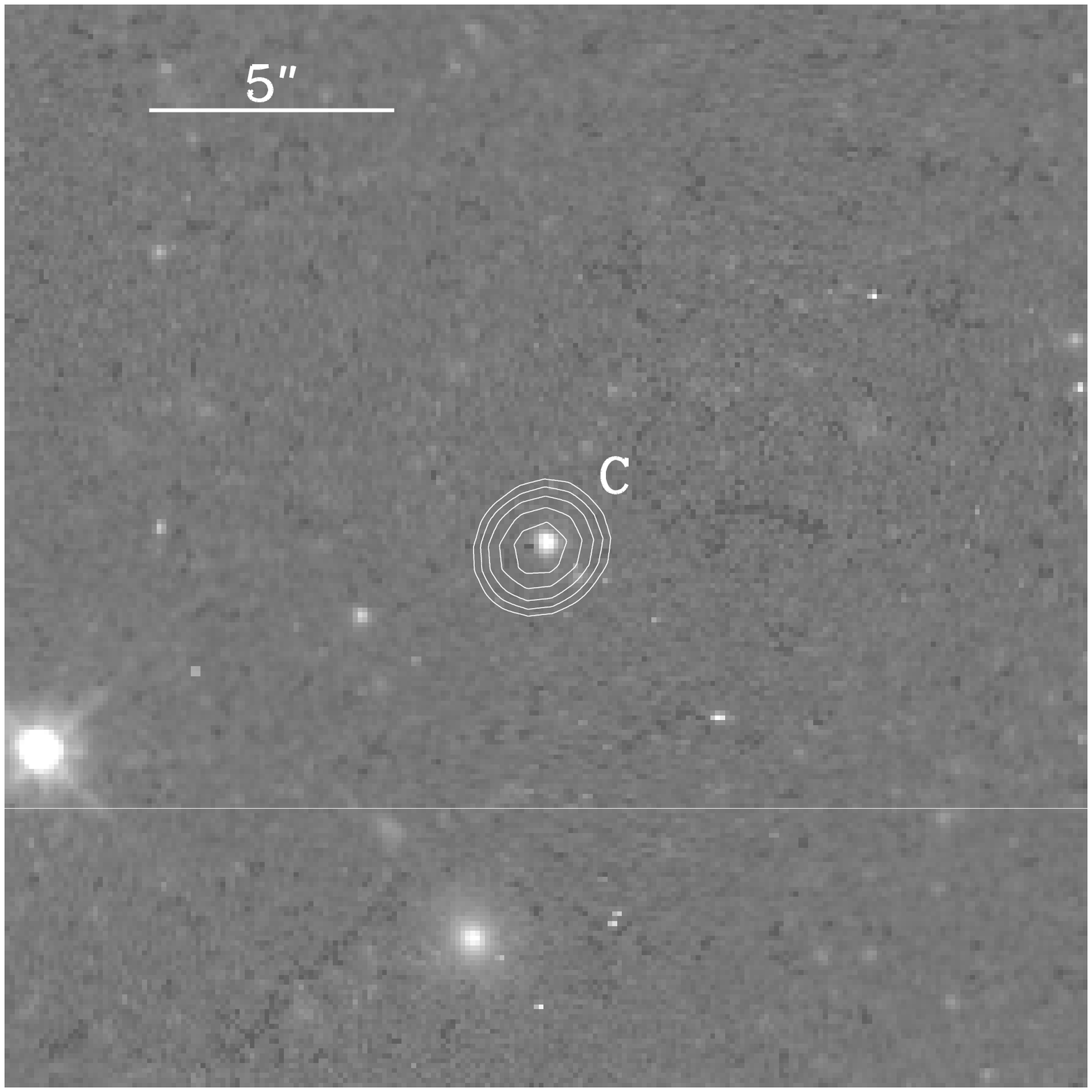}
\includegraphics[scale=0.25]{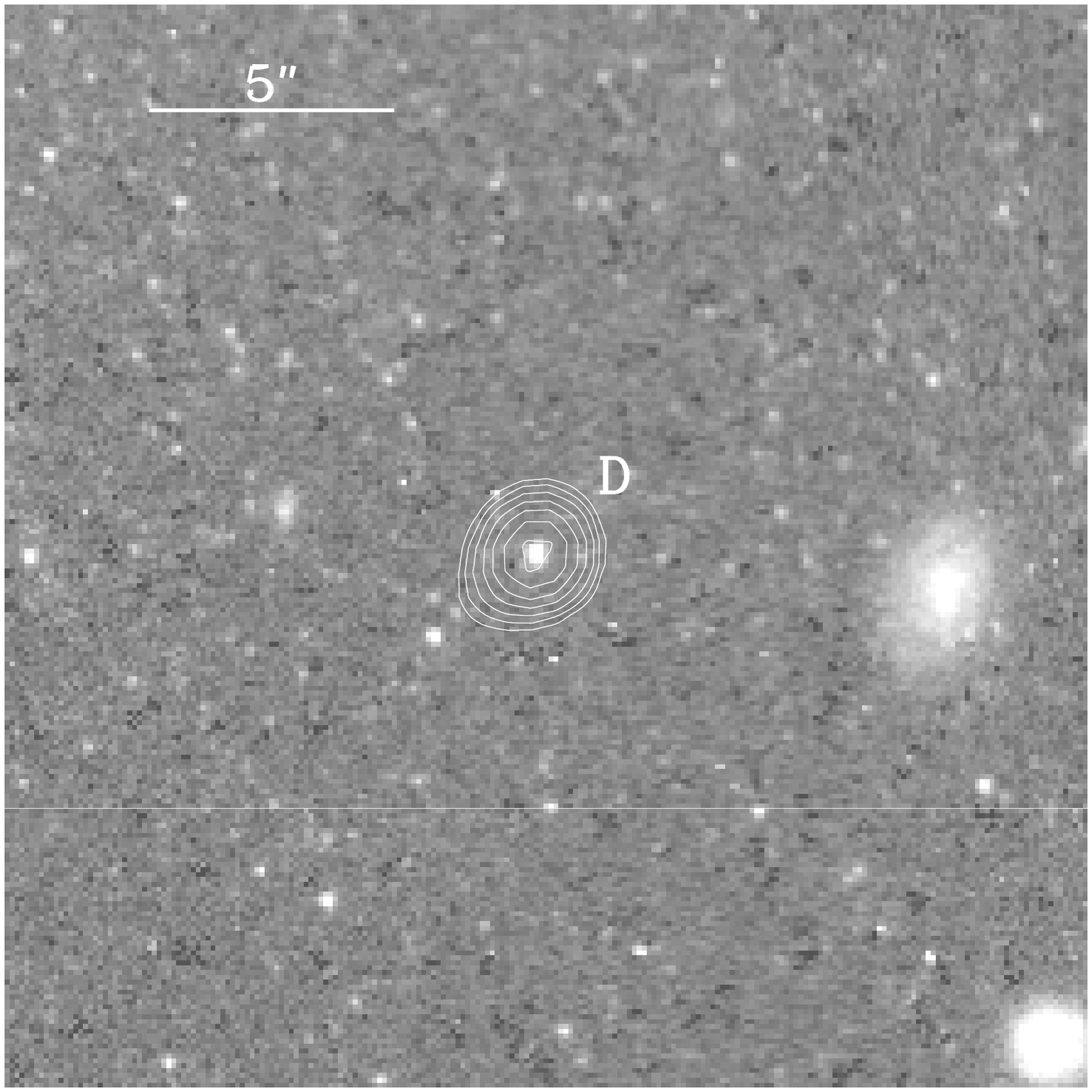}
\end{minipage}
\caption{Left: JKT grey-scale R band image of NGC6503 with the position of
the detected discrete sources marked with circles. Right: Detail of
(from top to bottom) the nuclear region, source C and source D, with
overlaid X-ray contours. The grey-scale images correspond to HST WFPC2
observations (PIs: Davis, Illingworth) obtained with the filters F606W
(for the nuclear image) and F814W (for the background objects - see
section 5.3).}
\end{figure*}

\begin{table*}
\caption{Detected discrete sources in NGC6503. Background subtracted
source counts ($SC$) in the 0.3-10 keV range are given for each
source. The significance of the detections was determined as
$SC/\sigma_{B}$, where $\sigma_{B}$ is the standard deviation of the
background and was computed as $1 + \sqrt{\rm{Background\ counts} +
0.75}$. This correction to the standard deviation gives a more
appropriate estimate of Poissonian errors for cases of low number
counts (Gehrels 1986). {\it Observed\/} fluxes are in units of
$10^{-14}$ ergs s$^{-1}$ cm$^{-2}$. {\it Intrinsic\/} luminosities and
fluxes (for those sources not associated with NGC6503) are in units of
ergs s$^{-1}$ and ergs s$^{-1}$ cm$^{-2}$, respectively. Hydrogen
column densities are in units of $10^{21}$ cm$^{-2}$. $a$: Results
from spectral fitting for source 6 (see text).}  \centering
\begin{tabular}{lllcrccccl} \hline
Source & \multicolumn{2}{c}{Position (J2000)} & Counts & Signif. & $\Gamma$ & $N_{H}$
& F$^{\rm obs}_{\rm x}$ & L$^{\rm int}_{\rm x}$--F$^{\rm int}_{\rm x}$ & Comments\\ \hline
1 & 17 49 12.48  &70 9 31.4&  $136 \pm 17$   	&65.2	&3.0	&1.0	&4.69	&$3.35\times10^{+38}$ &\\
2 & 17 49 26.43  &70 8 39.7&  $13  \pm 5$ 	&6.0	&1.5	&1.0	&0.90	&$5.85\times10^{+37}$ & NGC6503 Nucleus\\
3 & 17 49 27.89  &70 8 36.7&  $34  \pm 7$ 	&16.1	&1.5	&1.0	&2.34	&$1.52\times10^{+38}$ &\\
4 & 17 49 28.84  &70 8 32.7&  $34  \pm 7$ 	&16.1	&1.5	&1.0	&2.34	&$1.52\times10^{+38}$ & Opt ID: GC in NGC6503\\
5 & 17 49 29.09  &70 8 44.2&  $154 \pm 13$   	&73.8	&1.5	&1.0	&10.6	&$6.89\times10^{+38}$ &\\
6 & 17 49 31.72  &70 8 20.3&  $255 \pm 17$   	&122.4	&1.5	&5.0	&25.8	&$1.98\times10^{+39}$ & ULX in NGC6503\\
6$^{a}$&         &         &                 	&	&$kT=4$ keV&5.5	&17.4	&$1.46\times10^{+39}$ &\\
7 & 17 49 38.09  &70 8 31.9&  $7   \pm 4$ 	&3.2	&1.5	&1.0	&0.48	&$3.12\times10^{+37}$ &\\
A & 17 49 09.84  &70 7 26.4&  $9   \pm 4$ 	&4.1	&2.0	&0.4	&0.62	&$6.97\times10^{-15}$ & Opt ID: Bkg QSO\\
B & 17 49 11.26  &70 6 24.1&  $8   \pm 4$ 	&3.6	&2.0	&0.4	&0.55	&$6.19\times10^{-15}$ & Opt ID: Bkg QSO\\
C & 17 49 18.58  &70 6 31.8&  $24  \pm 6$ 	&11.3	&2.0	&0.4	&0.83	&$1.03\times10^{-14}$ & Opt ID: Bkg QSO\\
D & 17 49 26.65  &70 6 51.8&  $31  \pm 7$ 	&14.7	&2.0	&0.4	&2.14	&$2.41\times10^{-14}$ & Opt ID: Bkg QSO\\
\hline
\end{tabular}
\end{table*}

\section{Chandra observations}

X-ray observation of NGC6503 were obtained on the 23$^{rd}$ of March
2000 using the ACIS S3 chip onboard the Chandra satellite. The total
exposure time was 13 ksec and no flares of high background radiation
were observed during this period. The analysis of the X-ray data was
limited to a $\sim 6\arcmin \times 6\arcmin$ region centered on
NGC6503 which is coincident with a R-band image of the galaxy obtained
with the Jacobus Kapteyn Telescope (JKT) telescope (Johnson etal
2002). Although the ACIS sensitivity extends below 0.1 keV, the
spectral analysis of the observations was limited to the 0.3-10 keV
band due to poor calibration of the instrument response matrices below
this range.

\section{Diffuse emission}

Initial examination of the data revealed several discrete sources
coincident with the position of NGC6503, but no evidence for strong
extended emission associated with the galaxy. To quantify this we
measured the number of counts per pixels within concentric annuli
centred on the nucleus of the galaxy after discarding the discrete
sources. The result showed a clear but modest growth of the surface
brightness in the central region of the galaxy, from that of the
background level at a radius of $\sim 90\arcsec$ to an increment of a
$\sim 50\%$ above the background emission in the inner $\sim
40\arcsec$. The count rate associated with this diffuse component
corresponds to $\sim 0.01$ cps, or an unabsorbed flux of $\sim
6.4\times10^{-13}$ ergs s$^{-1}$ cm$^{-2}$ in the 0.3-10 keV energy
band, assuming a thermal model with $kT=5$ and a Galactic absorbing
column of $4\times10^{20}$ cm$^{-2}$. At a distance of 7Mpc the
luminosity of this component corresponds to $\sim 4\times10^{38}$ ergs
s$^{-1}$.

From our analysis of the ROSAT HRI data we reported a total flux of
$\sim 4.5\times10^{-13}$ ergs s$^{-1}$ cm$^{-2}$ in the 0.1-2.4 keV
energy band within a radius of 30\arcsec\, assuming the same spectral
parameters and absorbing column as above. This emission was
interpreted as an extended component associated with the central region
of the galaxy. 

About half of the flux from the diffuse component detected with
Chandra is expected in the ROSAT band, ie $\sim 0.8\times10^{-13}$
ergs s$^{-1}$ cm$^{-2}$, which corresponds to $\sim 17\%$ of the total
flux measured with ROSAT. Therefore the emission detected by the ROSAT
observations is dominated by the unresolved point sources and not by
the diffuse component seen with Chandra. The Chandra extended emission
could correspond to an unresolved population of low luminosity X-ray
sources, to truly diffuse emission from the hot ISM, or both.


\section{Compact Sources}

A source detection algorithm showed the presence of several discrete
X-ray sources in the NGC6503 Chandra data. For each discrete source a
background subtracted number of counts was obtained from within a $r =
3$ pixel aperture ($\sim 1.5$ arcsec). This encircles $\sim 90\%$ of
the photons at 1 keV. The small apertures used imply that on average
only $\sim 0.6$ counts are expected to correspond to the background
($\sim 0.8$ over the diffuse component in the central
region of the galaxy). The background was determined from an annular
region around the galaxy devoid of obvious X-ray sources. The final
list of sources was compiled from those sources with a significance
larger than 3 (see caption of Table 1). The determined counts and
associated significance can be found in Table 1. Seven sources were
found to be associated with NGC6503 (sources 1 to 7), while 4 sources
more probably correspond to background objects (sources A to
D). Figure 1 shows the JKT observation of NGC6503 with the position of
the sources labeled with increasing right ascension.

The astrometric accuracy of Chandra is observed to be better than
$\sim 0.6\arcsec$ \footnote{see the {\it Chandra\/} Proposers'
Observatory Guide v3.0, December 2000,
http://asc.harvard.edu/udocs/docs/POG/MPOG/}. We found that the
position of source 2 see in Figure 1 is in excellent agreement with the
optical nucleus of the galaxy. Also, faint optical counterparts are
found for sources C and D (see right panel of Figure 1).

The spatial extent of the sources was also studied. Spatial profiles
were determined by measuring the number of counts in 5 concentric
annuli of widths corresponding to 1 pixel. The distribution was then
compared with a generated PSF profile corresponding to a point source
observed at the same detector position. No significant deviations from
a point source was observed for the detected sources. From the model
PSF FWHM ($\sim 1.1-1.2$ pixels) an upper limit for linear size of the
X-ray emitting regions of $\sim 20$ pc is found for those sources
associated with NGC6503. Only source 5, and maybe source 6, show a
marginal spatial extent larger than that of a point source (see Figure
2). For these sources we measure a FWHM $\sim 1.3$ pixels.

To convert the observed count rates to fluxes we have used a
colour-colour diagram to infer the values to adopt for the source
intrinsic spectral shapes and intervening absorbing columns. Figure 3
shows the diagram using the observed count ratios in the 0.3-1, 1-2
and 2-7 keV energy ranges for the brightest sources associated with
NGC6503. The diagram shows that most of the sources show a hard
spectral shape, which could be described with a powerlaw of index
$\Gamma \sim 1.5$. Source 1, presents a slightly softer spectrum
characterised by $\Gamma \sim 2.5$ or thermal emission with $kT \sim 3$
keV. Although the column density is not well constrained, most sources
show evidence for some intrinsic absorption above the Galactic value,
while source 6 shows definite evidence for strong absorption.

Fluxes and luminosities presented in Table 1 have been obtained
assuming a spectral shape parameterised by a powerlaw with $\Gamma =
1.5$ and an absorbing column of $1\times10^{21}$ cm$^{-2}$ for all
sources found in NGC6503, except for source 1 which was parameterised
using an index $\Gamma = 2.5$, and source 6 where a $5\times10^{21}$
cm$^{-2}$ column density was adopted. Sources outside NGC6503 were
parameterised assuming a Galactic column density and a power law with
$\Gamma = 2.0$, typical of AGN.

We fitted the spectrum of source 6, which has the highest number of
counts, in order to check our results from the colour-colour diagram.
A single thermal (Mekal) component and a single PowerLaw (PL)
component were tried. In each case freezing the absorbing neutral
hydrogen column to the Galactic value resulted in unacceptable fits,
with $\chi^{2}_{\rm red} \ga 3$. Allowing for a free value of N$_{H}$
gave $\chi^{2}_{\rm red} \sim 1.3$ for the thermal model, while a
poorer fit was obtained for the power law. A double (M+PL) model did
not improve the quality of the fit. Finally, a double Mekal (M+M)
model gave $\chi^{2}_{\rm red} \sim 1.2$ for 12 degrees of freedom.

The fitted parameters for the single M model were N$_{H} =
5.5^{+2.1}_{-1.5}\times10^{21}$ cm$^{-2}$, $kT = 3.9^{+2.3}_{-1.2}$
keV. The M+M model gave N$_{H} = 1.0^{+3.8}_{-4.7}\times10^{22}$
cm$^{-2}$, $kT_{1} = 0.9^{+0.8}_{-0.2}$ keV, and $kT_{2} \sim 4$ keV
(no formal constraints were obtained for this parameter). These values
are in good agreement with those inferred from the colour-colour
diagram which shows that source 6 is characterised by hard and heavily
absorbed emission. Table 1 shows the flux and luminosity obtained from
the single M model. The derived flux remained constant when changing
from a single to a double Mekal component, but the luminosity
increased by $\sim 50\%$ in the M+M case.

The adoption of different power law indices to those used in Table 1
(keeping $\Gamma$ in the $1-3$ range) implied changes in the source
fluxes of at most $\sim 50\%$. The derived intrinsic emission is much
more sensitive to the adopted absorbing columns, particularly for the
softer sources. The statistical errors associated with source 4, for
example, are consistent with $\Gamma \sim 3.0$ and $N_{H} \sim
5\times10^{21}$ cm$^{-2}$ (see Figure 3). In this case the intrinsic
luminosity of the source would be $\sim 2.5$ times larger than that
quoted in Table 1. The presence of more complex spectral shapes could
add to the uncertainty of the derived values.

\begin{figure}
\centering
\includegraphics[angle=270,scale=0.4]{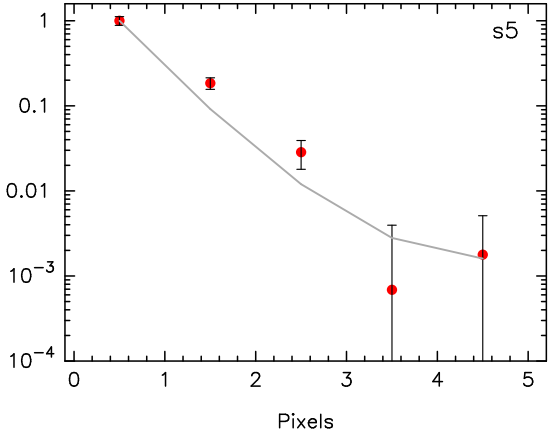}
\caption{Spatial profile of source 5 obtained from a full 0.3-10 keV
band image and compared with the model PSF (continuous lines). The
error bars correspond to 1$\sigma$ confidence levels.}
\end{figure}

\begin{figure} 
\centering
\includegraphics[angle=270,scale=0.48]{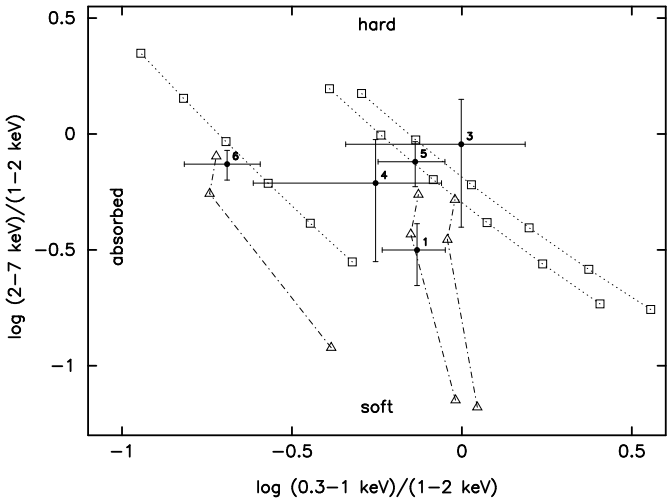}
\caption{Colour-colour diagram for X-ray discrete sources. 
The X-ray colours were defined as the ratio of counts observed in the
0.3-1.0 keV, 1.0-2.0 keV and 2.0-7.0 keV energy bands. One $\sigma$
error bars are shown. The grids of models correspond to single
Raymond-Smith plasmas (triangles - dash-dotted line) with temperatures
(from top to bottom) 6.0, 3.0, and 1.0 keV, and single power laws
(squares - dotted line) with index $\Gamma$ (from top to bottom) 0.5,
1.0, 1.5, 2.0, 2.5 and 3.0. From right to left successive grids
correspond to absorbing columns of $5\times10^{20}$ (just above Galactic),
$1\times10^{21}$ and $5\times10^{21}$ cm$^{-2}$ respectively.}
\end{figure}


\section{The source population in NGC6503}

Seven discrete sources with L$_{\rm x} \sim 10^{37-39}$ ergs s$^{-1}$
have been found associated with NGC6503. Sources 1, 2, 3, 4 and 7 are
within the luminosity range of classic X-ray binaries (L$_{\rm x} \la
10^{38}$ ergs s$^{-1}$ - SMC X-1, the brightest, persistent HMXRB in
the local group has a mean L$_{\rm x} \sim 2.2\times10^{38}$ ergs
s$^{-1}$, Helfand \& Moran 2001). Sources 5 and 6 have luminosities
well in excess of this value. As discussed in the previous section,
there can be large uncertainties associated with the adoption of
spectral parameters for the sources, not only due to the formal errors
in the colour ratios shown in Figure 3, but also due to the difficulty
in adopting an accurate description of more complex spectra. However,
sources 5 and 6 are the brightest sources in the field and, therefore,
their spectral shapes are fairly well constrained in our colour-colour
diagram. Also, both sources present a hard spectrum, and therefore
their derived intrinsic luminosities are less affected by the adopted
absorbing hydrogen column.

Sources 5 and 6 also showed marginal evidence for an extended spatial
profile, as discussed in the previous section. This, together with the
high L$_{\rm x}$, suggests the presence of multiple unresolved sources
and/or compact diffuse emission (eg, bubbles from SNRs). However, the
observed variability of source 6 (see below) implies that its
emission is dominated by a single powerful X-ray source.

\subsection{The ULX}

The spectral analysis of source 6 has provided clear evidence of heavy
obscuration and of an intrinsically hard spectrum. Its luminosity
(L$_{\rm x} \ga 10^{39}$ ergs s$^{-1}$) puts it within the range of
the Ultra Luminous Sources (ULX). The nature of these sources is not
yet well understood, but current evidence suggests that they are most
probably powered by accretion onto compact sources and are normally
associated with regions of active star formation (Roberts etal
2002). This also seems to be the case for source 6, which is nearly
coincident with a distinctive knot of blue emission ($B = 20.3;
U-B = -0.8$) seen in the disk of NGC6503, implying its possible
association with a star-forming region.

We have overlaid the Chandra and ROSAT HRI data for NGC6503 using as a
reference the positions of sources 1 and D, and applying the
corresponding scaling to account for the different plate scales. The
ROSAT observations had an integration time of $\sim 15$ ksec and were
obtained on the 8$^{th}$ of March 1994, ie approximately 6 years
before the Chandra observations. The instrument had a PSF FWHM $\sim
5\arcsec$ and a pixel size $\sim 0.5\arcsec$. Figure 4 shows the ROSAT
and Chandra observations of the central region of NGC6503 as X-ray
isocontours.  The position of the `nuclear' emission seen in the HRI
data is in good agreement with the flux centroid of the emission from
sources 3, 4 and 5 in the Chandra data. The nuclear source, 2, also
seems to have a counterpart in the ROSAT data, although this is well
within the noise level of the ROSAT observations. No significant
emission is seen in the ROSAT data at the position of source 6, the
brightest source seen during the Chandra observations.

Using the apertures shown in Figure 4 ($r = 11.5\arcsec$) we have
obtained an estimate of the observed ROSAT fluxes associated with the
`nuclear' source (ie, the integrated emission from sources 3, 4 and 5)
and source 6. From the `nuclear' aperture a total of $15.8 \pm 6.0$
counts were measured.  Assuming the spectral shape described in the
previous section this translates into an observed flux of
$4\times10^{-14}$ ergs s$^{-1}$ cm$^{-2}$ in the ROSAT 0.1-2.4 keV
energy band. This value is in good agreement with the Chandra flux of
$\sim 6\times10^{-14}$ ergs s$^{-1}$ cm$^{-2}$ from sources 3, 4 and 5
in the same energy range.

Source 6 was not significantly detected in the ROSAT data and a 90\%
CL upper limit of 6.6 counts was found adopting Bayesian statistics.
Assuming the same spectral shape as in Table 1, this corresponds to an
observed flux of $2\times10^{-14}$ ergs s$^{-1}$ cm$^{-2}$ in the
ROSAT 0.1-2.4 keV energy band. The observed Chandra flux for source 6
in the ROSAT band corresponds to $6\times10^{-14}$ ergs s$^{-1}$
cm$^{-2}$, ie, a change in luminosity by at least a factor 3 in 6
years for this ULX.

Long-term variability has already been observed in the other ULXs.
NGC4559 (X-1), for example, has shown a decrease in its luminosity by
a factor of $\sim 2$ in the last 10 years (Roberts etal 2002). The
variability of S6 in NGC6503 seems to be more extreme, since the
ROSAT flux is a strict upper limit. 

Another interesting characteristic of source 6 is the large observed
absorbing column of N$_{H} \sim 5\times10^{21}$ cm$^{-2}$. This is not
atypical amongst other ULXs, many of which show columns of a few
$10^{21}$ cm$^{-2}$, sometimes in excess of the extinction derived for
the parent galaxy (Roberts etal 2002). It is not clear whether this
extra absorption is inherent to the ULXs or a consequence of their
association with dusty star forming regions.

\begin{figure}
\centering
\includegraphics[scale=0.45]{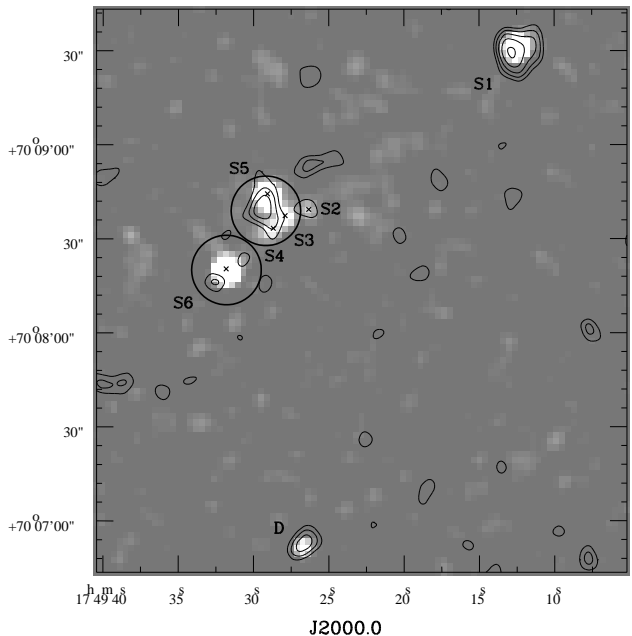}
\caption{X-ray emission associated with NGC6503: ROSAT HRI contours
are displayed onto a gray scale image of the Chandra observations. The
HRI data have been binned into 2\arcsec\ pixels in order to increase
the signal-to-noise ratio. The resolution of the Chandra data has been
degraded to that of the HRI. Both images have been slightly smoothed
with a Gaussian distribution. Sources 1 and D have been used to
register the data. The apertures used to extract counts from the ROSAT
data are shown as black circles. The positions of Chandra sources
detected in the central part of the galaxy are shown with crosses.}
\end{figure}

\subsection{The nuclear source}

A weak, compact X-ray source is associated with the nucleus of
NGC6503. It presents a linear size $\sim 20$ pc and a luminosity of
$\sim 4\times10^{37}$ ergs s$^{-1}$ and $\sim 6\times10^{37}$ ergs
s$^{-1}$ in the 2-10 and 0.3-10 keV energy bands, respectively. With a
nuclear H$\alpha$ luminosity $\sim 4\times 10^{37}$ ergs s$^{-1}$ (Ho,
Filippenko \& Sargent 1995) a low luminosity AGN in NGC6503 should
have a 2-10 keV L$_{\rm x}$ about a factor 5 larger (Ho etal 2001),
unless the source is significantly obscured (notice, however, that
NGC6503 lies at the very faint end of the L$_{\rm H\alpha}$-L$_{\rm
x}$ relationship determined by Ho etal, where the correlation is less
certain).  The nuclear source in NGC6503 is, however, soft, with only
4 counts observed above 2 keV. This, together with the Liner-starburst
classification by Lira etal (2002) based on optical spectroscopy,
makes the presence of a low luminosity AGN very unlikely.

The nuclear X-ray luminosity in NGC6503 is well within the range of
individual sources seen in our galaxy, such as X-ray binaries and
luminous Supernova Remnants. But there is also the possibility that
the nuclear source is in fact formed by a group of less luminous
objects. As an example, ROSAT observations of M31 detected a compact
source associated with the nucleus at the instrument resolution of
$\sim 5\arcsec$ ($\sim 20$ pc at the distance of M31; Primini, Forman
\& Jones 1993). Its luminosity in the ROSAT band was found to be $\sim
2\times10^{37}$ ergs s$^{-1}$. Subsequent Chandra observations with a
resolution of $\sim 2$ pc have shown that the source is in fact formed
by 5 distinctive objects (Garcia etal 2001).

\subsection{Other sources in NGC6503}

We have used archive HST WFPC F606W observations (PI: Illingworth) of
the nuclear region of NGC6503 to search for optical counterparts to
the X-ray sources (see right panel of Figure 1). Apart from the
nuclear source, 2, a faint optical counterpart is also seen for source
3. Aperture photometry gives a magnitude of 22.5 in the $V_{KC}$ band
filter, corresponding to an absolute magnitude of -6.7. The object is
resolved with a FWHM $\sim 0.25\arcsec$ or a linear size of 8.5 pc at
a distance of 7 Mpc. These parameters suggest that the optical
counterpart is a globular cluster in NGC6503, with a luminosity $\sim
1$ magnitude fainter than the peak value seen in the luminosity
distribution function of globular clusters in our Galaxy (Harris
2001).

Studies of the Milky Way and M31 show that bright X-ray sources are
often associated with globular clusters. The population of globular
clusters in M31 is considerably more luminous than that seen in our
Galaxy (Di Stefano etal 2002). The brightest globular cluster detected
in M31, Bo~375, has an X-ray luminosity in the Chandra band similar to
source 4 found in NGC6503, $\ga 10^{38}$ ergs s$^{-1}$. Therefore, if
our identification of source 4 as a globular cluster in NGC6503 is
correct, this source could be representative of the bright end of the
X-ray luminosity distribution of globular clusters in that galaxy.


\section{The background sources}

Four background sources were detected in the field of view of
NGC6503. Two of them (C and D) had faint optical counterparts in our
JKT R-band image. We retrieved HST WFPC2 centred in the sky region
where the sources were located (PI: Davies). All sources were clearly
identified with optical counterparts of magnitudes 24.6, 25.6, 21.7
and 22.3 for A, B, C and D respectively, in the F814W filter (and
transformed to $I_{C}$ magnitudes). The right panel of Figure 1 shows
the two brightest WFPC counterparts.

Assuming, as before, a power law X-ray spectrum with $\Gamma = 2$,
these sources will have $\sim 45\%$ of their 0.3-10 keV flux given in
Table 1 in the 2-10 keV band. The optical and X-ray fluxes are in very
good agreement with those determined for the sources found in the
Chandra deep survey of the HDF-N (Hornschemeier etal
2001). Spectroscopic follow up of these sources has showed that the
majority correspond to AGNs at redshifts $\la 1$ (Hornschemeier etal
2001, Tozzi etal 2001).


\section{Summary}

We have obtained high resolution Chandra X-ray observations of the
nearby spiral galaxy NGC6503. The data have shown the presence of
faint diffuse emission around the central region of the galaxy with a
luminosity of $\sim 10^{39}$ ergs s$^{-1}$ in the 0.3-10 keV energy
range. A total of 11 compact sources were identified in the field of
view, 7 of which seem to be associated with NGC6503. The remaining
sources seem to correspond to background quasars. 

A weak, L$_{\rm x} \sim 6\times10^{37}$ ergs s$^{-1}$, source is found
to be coincident with the Liner-starburst nucleus of the
galaxy. Another X-ray source has a faint optical counterpart which
might correspond to a globular cluster in NGC6503. Finally, a L$_{\rm
x} \ga 10^{39}$ ergs s$^{-1}$ ULX has been identified which is nearly
coincident with a distinctive knot of blue emission seen in the disk
of the galaxy. Previous ROSAT observations obtained 6 years before
show that this source has varied by at least a factor 3.


\bibliographystyle{}

\end{document}